\documentclass[sigconf,nonacm]{acmart}

\usepackage{listings}
\usepackage{xcolor}
\usepackage[bottom]{footmisc}

\colorlet{punct}{red!60!black}
\definecolor{background}{HTML}{EEEEEE}
\definecolor{delim}{RGB}{20,105,176}
\colorlet{numb}{magenta!60!black}

\lstdefinelanguage{json}{
    basicstyle=\normalfont\ttfamily,
    numbers=left,
    numberstyle=\scriptsize,
    stepnumber=1,
    numbersep=8pt,
    showstringspaces=false,
    breaklines=true,
    frame=lines,
    backgroundcolor=\color{background},
    literate=
     *{0}{{{\color{numb}0}}}{1}
      {1}{{{\color{numb}1}}}{1}
      {2}{{{\color{numb}2}}}{1}
      {3}{{{\color{numb}3}}}{1}
      {4}{{{\color{numb}4}}}{1}
      {5}{{{\color{numb}5}}}{1}
      {6}{{{\color{numb}6}}}{1}
      {7}{{{\color{numb}7}}}{1}
      {8}{{{\color{numb}8}}}{1}
      {9}{{{\color{numb}9}}}{1}
      {:}{{{\color{punct}{:}}}}{1}
      {,}{{{\color{punct}{,}}}}{1}
      {\{}{{{\color{delim}{\{}}}}{1}
      {\}}{{{\color{delim}{\}}}}}{1}
      {[}{{{\color{delim}{[}}}}{1}
      {]}{{{\color{delim}{]}}}}{1},
}

\AtBeginDocument{%
  \providecommand\BibTeX{{%
    \normalfont B\kern-0.5em{\scshape i\kern-0.25em b}\kern-0.8em\TeX}}}

\copyrightyear{2021} 
\acmYear{2021} 
\setcopyright{acmlicensed}
\acmConference[CHIIR '21]{Proceedings of the 2021 ACM SIGIR Conference on Human Information Interaction and Retrieval}{March 14--19, 2021}{Canberra, ACT, Australia}
\acmBooktitle{Proceedings of the 2021 ACM SIGIR Conference on Human Information Interaction and Retrieval (CHIIR '21), March 14--19, 2021, Canberra, ACT, Australia}
\acmPrice{15.00}
\acmDOI{10.1145/3406522.3446043}
\acmISBN{978-1-4503-8055-3/21/03}

\begin{document}
\fancyhead{} 

\title{Dataset of Natural Language Queries for E-Commerce}


\author{\mbox{Andrea Papenmeier, Dagmar Kern, Daniel Hienert}}
\email{firstname.lastname@gesis.org}
\affiliation{%
  \institution{GESIS – Leibniz Institute for the Social Sciences}
  \city{Cologne}
  \country{Germany}
}

\author{Alfred Sliwa, Ahmet Aker, Norbert Fuhr}
\email{firstname.lastname@uni-due.de}
\affiliation{%
  \institution{University of Duisburg-Essen}
  \city{Duisburg}
  \country{Germany}
}

\renewcommand{\shortauthors}{Papenmeier et al.}

\begin{abstract}
Shopping online is more and more frequent in our everyday life. For e-commerce search systems, understanding natural language coming through voice assistants, chatbots or from conversational search is an essential ability to understand what the user really wants. However, evaluation datasets with natural and detailed information needs of product-seekers which could be used for research do not exist. Due to privacy issues and competitive consequences, only few datasets with real user search queries from logs are openly available. In this paper, we present a dataset of 3,540 natural language queries in two domains that describe what users want when searching for a laptop or a jacket of their choice. The dataset contains annotations of vague terms and key facts of 1,754 laptop queries. This dataset opens up a range of research opportunities in the fields of natural language processing and (interactive) information retrieval for product search.
\end{abstract}

\ccsdesc[300]{Human-centered computing~Human computer interaction (HCI)}
\ccsdesc[500]{Human-centered computing~Natural language interfaces}
\ccsdesc[500]{Information systems~Query intent}
\ccsdesc[300]{Computing methodologies~Discourse, dialogue and pragmatics}

\keywords{Dataset; Natural Language Query; User Intent; E-Commerce.}

\maketitle

\section{Introduction}
To search for products online is an everyday activity of millions of users, with the market share of e-commerce continually increasing \cite{uscommerce2020}. 
Understanding natural language in deep is the upcoming key technology for searching in e-commerce and the general Web: Voice assistants rely on processing spoken natural language, chatbots need to extract the user's information need from written natural language, and the research field of conversational search explores a dialogue-driven approach to support finding the right information. Research on query formulation with children shows that natural language is the intuitive way of interacting with search engines \cite{kammerer2012children}. Keyword search, however, is an artefact of the search engine's inability to process open vocabularies and extract the essential key facts from a natural language text. 
Research on query and information need formulation has mainly built on log data \cite{sadikov2010clustering,shen2011sparse,yan2017building} or proprietary data \cite{guy2016searching}. Log data is not suitable to investigate the natural information need, as the system influences the user, i.e. if users believe the system can only handle keywords, they formulate their query accordingly \cite{kato2014cognitive}. Some smaller datasets of natural language information needs exist, e.g. as collected by Kato et al. \cite{kato2014cognitive}. In the book domain, The CLEF Social Book Search dataset \cite{koolen2016overview} provides 120 natural language information needs. The data originates from an online discussion forum, which represents a non-controlled data collection setting.
In previous research on a small dataset of 132 natural language queries of laptops, we have shown that natural language queries have the potential to reveal more information about the user's target product than queries issued on current search engines \cite{papenmeier2020modern}. However, available datasets are not big enough to train automated systems to process natural language information needs.

In this work, we collected and curated a large dataset containing 3,540 natural language queries for two product domains (laptops and jackets). Unlike existing datasets, our product queries were collected in a controlled experiment from participants with a broad range of domain knowledge. For the laptop domain and a subset of the jacket domain, we offer manual annotations of the key product facts mentioned in the descriptions, and vague words contained in the texts. 
With this dataset, we contribute a valuable resource for the field of natural language processing and interactive information retrieval in the context of product search.

\section{Related Work}
\subsection{Data Collections of Information Needs}
One of the most chosen instruments for understanding user needs is the extraction from \textit{query logs}. For example, a rough distinction is made by Broder \cite{broder2002} who defined the user intent of web search queries to be either informational, navigational or transactional. Several research works cluster either search queries from logs \cite{sadikov2010clustering,shen2011sparse,yan2017building} or transcribed voice logs \cite{guy2016searching} into groups of user intent. There is certainly a big difference between the user's natural information need and the short keyword-like queries which can be found in query logs. A more detailed description of information needs is given in \textit{TREC} \cite{vorhees2005} topics and the works based upon (e.g. \cite{Bailey2015,potthast2013exploratory}). Here, the situation and context of the search intent are described in more detail and formed into background stories or search tasks. The CLEF Social Book Search dataset \cite{koolen2016overview} contains a collection of 120 natural language information needs of books extracted from an online discussion forum. Bogers et al. \cite{bogers2019what} likewise focus on forums to extract natural language information needs. They annotated 1041 information needs (503 in the book domain, 538 in the movies domain) with respect to ``relevance aspects'', i.e., requirements for the search target.

Another direction is \textit{conversational search}, where the conversation approach fetches more details on the actual information need of the user. Evaluation datasets base on already existing question-answering interactions as available in forums or dialogue support systems. For example, MS Dialog \cite{Qu2018} consists of 2,400 annotated dialogues about Microsoft products, and Penha et al. \cite{penha2019introducing} provide a dataset of 80k conversations across 14 domains which they extracted from Stack Exchange. 

\textit{Human-human conversations} in real as well as in Wizard-of-Oz situations in which humans simulate the system are another source of naturally formulated information needs. For example, the CCPE-M dataset \cite{Radlinski2019} focuses on movie preferences, while the Frames dataset \cite{el2017frames} focus on the travelling domain with dialogues gathered in a human-human conversation using SLACK. An additional approach in conversational search is to ask \textit{clarifying questions}. The Qulac dataset \cite{Aliannejadi2019} collected 10k clarifying questions with answers for 198 TREC topics in a crowdsourcing experiment. 

Datasets on \textit{spoken information-seeking conversations} between humans provide audio, transcriptions and additional annotations. Spoken conversations show differences to written conversations and can be used for the evaluation of software agents such as Siri, Google Now, or Cortana. The MISC dataset \cite{thomas2017misc} contains audio, video, transcripts, affectual and physiological signals, computer use and survey data for five different search tasks based on topics from previous research. The SCSdata~\cite{trippas2017how} contains 101 transcribed conversations with annotations and video to solve information needs based on nine search tasks and background stories.

\subsection{Product Search in E-commerce}
\textit{Product search} and \textit{e-commerce} is a rather new field in academic research and has specialised challenges for information retrieval: documents, queries, relevance, ranking, recommenders, and user interactions are different from well-known research areas such as Web search \cite{tsagkias2020challenges}. On the level of \textit{user intents}, Su et al.~\cite{su2018} distinguish between three different user goals: target finding, decision making, and exploration, which all have different behaviour patterns of query formulation, browsing, and clicking. Sondi et al.~\cite{Sondhi2018} report on another taxonomy of queries generated by clustering queries from a log: shallow exploration, major-item shopping, targeted purchase, minor-item shopping, and hard-choice shopping. \textit{Conversational e-commerce} is a new area where the user conducts a dialogue with the conversational system by voice or chat to find the right product or to get help. The dialogue needs to be natural so that the customer feels engaged. Therefore, understanding the overall intent of the user's request is essential \cite{tsagkias2020challenges}.

A current paper on research in e-commerce from the SIGIR Forum \cite{tsagkias2020challenges} lists 28 datasets for e-commerce search and recommendations. However, most of the datasets focus on product catalogues and taxonomies as well as on reviews and recommendations. Only two datasets contain search logs with user queries. Most research works in the area of e-commerce and product search of global players use their own query logs to improve their own search systems (e.g. eBay \cite{Trotman2017}, Amazon \cite{sorokina2016amazon} or Alibaba \cite{wang2018billion}), but keep the logs confidential. However, even if publicly available, these datasets contain only keyword queries and not the natural information needs the user is thinking of before entering it into a search bar.

Although the use case of product search is essential in e-commerce, there is little data about the genuine information needs of product buyers. To fill the gap of an openly available and controlled collected research dataset, we present in this work a collection of 3,540 natural language queries in product search.

\section{Dataset Generation}

\subsection{Data Collection}
We recruited 1,818 participants on the crowdsourcing platform Prolific\footnote{https://www.prolific.co} to participate in the experiment. To avoid the effects of the individual language level on the formulations, participants had to be native English speakers. Furthermore, participants were not allowed to use a mobile phone to complete the survey in order to avoid effects from small screen and keyboard sizes. 

After giving informed consent, participants were either asked to describe a laptop (imagining their current laptop broke down recently) or a jacket (imagining they lost their jacket). We define those product descriptions as \textit{natural language queries}. All participants completed both tasks, but in randomised order. The task description and the full questionnaire are available online\footnote{\url{https://git.gesis.org/papenmaa/chiir21_naturallanguagequeries/tree/master/Questionnaire}}.
Finally, the participants answered questions about their demographic background: (1) their age, (2) their gender, and (3) their self-assessed domain knowledge for both domains (on a scale of 1 = ``no knowledge'' to 7 = ``high/expert knowledge''). Table \ref{tab:participants_statistics} presents the demographic characteristics of the participants for both domains.

All descriptions were manually filtered to eliminate queries that were invalid due to their form (e.g. empty strings) or their content (e.g. the text described a different product, the text did not describe any product, the text was a meta-comment of the participant about the task). From the 1,818 participants, we curated a dataset containing 1754 valid laptop queries and 1786 valid jacket queries.

\begin{table}[t]
	\centering
	\begin{tabular}{l|r|r} 
	    \toprule
	    & \textbf{Laptop} & \textbf{Jacket} \\ 
	    \midrule
	    Participants (total) &                 1818 & 1818 \\
	    Valid queries (total) &           1754 & 1786 \\
	    Age (mean, std) &                      36 (13) & 36 (13) \\
	    Gender distribution (m, f, d) &        700, 1040, 12 & 718, 1054, 12 \\ 
	    Domain knowledge (mean, std) &               4.5 (1.5) & 4.6 (1.3) \\ \bottomrule
	\end{tabular}
	\caption{Participants statistics for the laptop and jacket queries dataset.}
	\label{tab:participants_statistics}
\end{table}

\subsection{Data Annotation}
After collecting the natural language queries, we recruited 20 annotators who were taking part in a seminar at our institution. Each laptop query was annotated by three annotators concerning key facts and vague words. Key facts are words or phrases describing requirements of the product, while vague words are words (within key facts) which are ambiguous and depending on interpretation. From the jacket corpus, so far, a subset of 363 queries was annotated concerning the key facts for cross-domain evaluation. 

Before starting the annotation task, annotators received the definition of key facts and vague words, examples, and guidance on how to handle negations and borderline cases for vagueness\footnote{The annotation material is available at \url{https://git.gesis.org/papenmaa/chiir21_naturallanguagequeries/blob/master/Annotationguideline.pdf}}. Annotators also discussed the guidelines in a plenary session. The annotation process was conducted on Doccano\footnote{https://github.com/doccano/doccano} as a sequence labelling task. Annotators labelled key facts (consisting of one or more words), e.g. (shown here in bold):
\begin{quote}
    \textit{I would buy a \textbf{basic} laptop of \textbf{any brand}, one with \textbf{good reviews}.}
\end{quote}
Furthermore, annotators labelled vague words, e.g.(shown here in bold):
\begin{quote}
    \textit{I would buy a \textbf{basic} laptop of any brand, one with \textbf{good} reviews.}
\end{quote}
The inter-annotator agreement (Krippendorff's alpha, on word-level) on the laptop queries is .697 for the annotation of key facts and .653 for the annotation of vague words. The annotations of jacket key facts have a mean inter-annotator agreement of .697. 
We provide the implementation of the agreement measure together with the dataset\footnote{Jupyter Notebook available at \url{https://git.gesis.org/papenmaa/chiir21_naturallanguagequeries/blob/master/IAA.ipynb}}. Table \ref{tab:dataset_statistics} shows detailed annotation statistics for the laptop domain. 

Listing 1 presents a single data point from the laptop domain of the final dataset, containing a unique ID, the domain, the original text (unprocessed), information about the user who wrote the text, and the annotations for both key facts and vague words. Each data point contains the individual annotations as well as a combined annotation showing words that were labelled by at least two annotators. The annotated words are identified by the word itself and its position (character-level offset) in the text. The annotated words of the key facts are also available as text segments, where the individual labels have been taken into account. 

\begin{table}[tb]
	\centering
	\begin{tabular}{l|r}
	    \toprule
	    & \textbf{Laptop} \\
	    \midrule
	    Queries (total) & 1754\\
	    Words per query (mean, std) & 35 (20)\\
	    
	    \textbf{Key facts} & \\
	    Annotated queries (total) & 1752\\
	    Annotated words per query (mean, std) & 10.6 (6.5)\\
	    Inter-annotator agreement & .697\\
	    
	    \textbf{Vague words} & \\
	    Annotated queries (total) & 1686\\
	    Annotated words per query (mean, std) & 3.2 (2.5) \\
	    Inter-annotator agreement & .653\\ \bottomrule
	\end{tabular}
	\caption{Dataset statistics for the laptop corpus.}
	\label{tab:dataset_statistics}
\end{table}

\subsection{Dataset Availability}
The dataset is publicly available\footnote{\url{https://git.gesis.org/papenmaa/chiir21_naturallanguagequeries}} under a CC BY-NC-SA 3.0 licence. The repository is hosted by GESIS -- Leibniz Institute for the Social Sciences, a well-known data provider for social science data. We provide the dataset in JSONL and CSV format, along with a description of variables and the annotation guidelines. Additionally, we provide a Jupyter Notebook with code to import the dataset into Python, perform basic statistical analyses, calculate the inter-annotator agreement, and access single data points. 
\newpage
\begin{lstlisting}[caption={Structure of a single data point in the data set.},language=json,firstnumber=1,basicstyle=\small]
{ "ID" : 1887,
  "domain" : 'laptop'
  "text" : 'I want a laptop primarily for internet use, it needs to be light with a long battery life.',
  "user" : {
    "age" : 47,
    "domain knowledge" : 3,
    "gender" : 'male'
  },
  "vague words" : {
    "annotator1" : [['light',59], ['long',72]],
    "annotator2" : [['light',59], ['long',72]],
    "annotator3" : [['light',59], ['long',72]],
    "annotation_by_2" : [['light',59], ['long',72]],
    "IAA" : 1.0
  },
  "key facts" : {
    "annotator1" : [['light',59], ['long',72], ['battery',77], ['life',85]],
    "annotator2" : [['internet',30], ['use',39], ['light',59], ['long',72], ['battery',77], ['life',85]],
    "annotator3" : [['light',59], ['long',72], ['battery',77], ['life',85]],
    "annotation_by_2" : [['light',59], ['long',72], ['battery',77], ['life',85]],
    "segments" : ['long battery life', 'light'],
    "IAA" : 0.8107
  }
}
\end{lstlisting}

\section{Use Cases}
The proposed natural language queries dataset can be leveraged for multiple tasks in the field of Natural Language Processing (NLP) and Information Retrieval (IR). The former could use this dataset to understand the semantic intents in natural language queries, while the latter could profit from building up (domain-specific) retrieval models for vague conditions. We discuss potential use cases in the following section.

\subsection{Natural Language Processing}

\subsubsection{Spelling correction}
The presented natural language queries have been written by users without any formal restrictions and thus contain many typographical and grammatical errors. The following example from our dataset demonstrates this issue: \textit{``i would  buy a lenovo as u can also use rhem as tablets which isvery handy''}. This example contains misspelled terms like \textit{``rhem''} and fusion error terms such as \textit{``isvery''}. Furthermore, entity-specific errors like misspelt brand names, e.g. \textit{``Lesovo''}, and domain-specific slang expressions like \textit{``has at least 8 gigs or ram''} are recurring phenomenons in this dataset. Containing different spelling error types and colloquial language expressions, this dataset calls for correction models in order to proceed with tasks like named entity recognition or information retrieval in product search (cf. \cite{wang2019multi}). The development of preprocessing techniques regarding raw natural language queries can be researched by investigating this dataset.

\subsubsection{Vagueness}
\label{subsubsec:vague}
One common problem in information retrieval is the vocabulary mismatch between the user's query language and the system's indexing language \cite{furnas1987vocabulary}. This is due to the vague information needs on the user-side where one is not able to sharpen the borders of different concepts, e.g.:
\begin{quote}
    \textit{I would like one with good battery and high RAM that boots relatively quickly}
\end{quote}
The vagueness problem can increase with a higher lack of domain knowledge \cite{papenmeier2020modern}. Hence, developing automatised models that are capable of recognising vague phrases in product search are needed.

This dataset is annotated on word-level regarding vague expressions. Classifying such vague words is helpful to distinguish between specific and vague conditions. In the case of specific conditions, a retrieval model can try to match such conditions with retailer-generated product information to filter the results. However, in case of vague query conditions, it is not straightforward to apply such filtering. Therefore, product retrieval systems could use user-generated content of products, e.g. user reviews, to filter for user requirements that are not entailed in the retailer-generated fields, e.g. product quality \cite{Moraes20}. In previous work, we demonstrated that user reviews are highly correlating with natural language queries according to lexical matching measurements \cite{papenmeier2020modern}.

\subsection{Attribute Mapping}
Natural language queries differ to keyword queries according to length, i.e. number of query terms, and the desired amount of conditions a certain object needs to meet. Named Entity Recognition (NER) is one task that aims to identify the different categories in natural language texts and can be applied to search queries \cite{jiafeng2009named}. As this dataset has been annotated on key fact-level, i.e. requirements that a product needs to satisfy, it can be used to research automatic matching of unstructured to structured information in product search (cf. \cite{tsagkias2020challenges}). Identifying the attribute domains of these key facts is useful for product retrieval systems as matching procedures can be conducted with the technical fields of products.

As natural language queries are characterised by a more complex structure opposed to keyword queries, another interesting task is to parse their semantic structure. Understanding and representing the meaning is more beneficial than using lexical matching methods like BM25.


\subsection{Product Query Classification}
This dataset has been annotated for the domain of laptops as well as for the domain of jackets. Product retrieval systems require category identification of a search query before applying the matching models. Misunderstanding the query's domain will lead to dissatisfying results. Product query classification has already been researched in the case of keyword search query data \cite{skinner2019commerce, yu2020query}. However, solving product query classification on natural language queries could initiate the investigation of more sophisticated algorithms.

\section{Conclusion and Future Work}
Natural language is an interesting challenge in product search. Currently, only few research work has focussed on collecting the unbiased natural information need of search engine users. We provide a dataset of 3,540 natural language queries of laptops and jackets. We annotated 1,754 laptop queries concerning the contained key facts and vague words, and 363 jacket queries.

As this dataset is part of an ongoing research project, we plan to enrich the dataset further. First, we aim to annotate the remainder of the jacket queries with vague words and key facts to have a fully comparable dataset in a second product domain. Secondly, we plan to add clean versions of the queries, corrected for spelling mistakes and punctuation. Thirdly, to enable work on interactive information retrieval and user experience design, the key facts should be matched to structured product attributes. In \cite{papenmeier2020modern}, we have made a first investigation of a smaller dataset (N = 132), where we mapped key facts to facets of existing product search engines and clustered unmatched key facts to determine new facets. To train classifiers on matching key facts to the correct facets, a ground truth is needed which we would like to add to the dataset in the future. Finally, for a subset of the product queries, we aim to add relevant products from a product pool to facilitate retrieval experiments.

For deep learning methods, however, the proposed dataset has a rather small sample size and could be further enlarged. Similarly to previous datasets, the proposed dataset is restricted to two product domains. To facilitate insights into the generalisability of models based on this dataset, more product domains should be added.

\begin{acks}
    This work was partly funded by the DFG, grant no. 388815326; the VACOS project at GESIS and the University of Duisburg-Essen.
\end{acks}

\balance
\bibliographystyle{ACM-Reference-Format}
\bibliography{chiirsp11-bib}


\begin{thebibliography}{33}


\ifx \showCODEN    \undefined \def \showCODEN     #1{\unskip}     \fi
\ifx \showDOI      \undefined \def \showDOI       #1{#1}\fi
\ifx \showISBNx    \undefined \def \showISBNx     #1{\unskip}     \fi
\ifx \showISBNxiii \undefined \def \showISBNxiii  #1{\unskip}     \fi
\ifx \showISSN     \undefined \def \showISSN      #1{\unskip}     \fi
\ifx \showLCCN     \undefined \def \showLCCN      #1{\unskip}     \fi
\ifx \shownote     \undefined \def \shownote      #1{#1}          \fi
\ifx \showarticletitle \undefined \def \showarticletitle #1{#1}   \fi
\ifx \showURL      \undefined \def \showURL       {\relax}        \fi
\providecommand\bibfield[2]{#2}
\providecommand\bibinfo[2]{#2}
\providecommand\natexlab[1]{#1}
\providecommand\showeprint[2][]{arXiv:#2}

\bibitem[\protect\citeauthoryear{Aliannejadi, Zamani, Crestani, and
  Croft}{Aliannejadi et~al\mbox{.}}{2019}]%
        {Aliannejadi2019}
\bibfield{author}{\bibinfo{person}{Mohammad Aliannejadi},
  \bibinfo{person}{Hamed Zamani}, \bibinfo{person}{Fabio Crestani}, {and}
  \bibinfo{person}{W.~Bruce Croft}.} \bibinfo{year}{2019}\natexlab{}.
\newblock \showarticletitle{Asking Clarifying Questions in Open-Domain
  Information-Seeking Conversations}. In \bibinfo{booktitle}{\emph{Proceedings
  of the 42nd International ACM SIGIR Conference on Research and Development in
  Information Retrieval}} (Paris, France) \emph{(\bibinfo{series}{SIGIR'19})}.
  \bibinfo{publisher}{Association for Computing Machinery},
  \bibinfo{address}{New York, NY, USA}, \bibinfo{pages}{475–484}.
\newblock
\showISBNx{9781450361729}
\urldef\tempurl%
\url{https://doi.org/10.1145/3331184.3331265}
\showDOI{\tempurl}


\bibitem[\protect\citeauthoryear{Bailey, Moffat, Scholer, and Thomas}{Bailey
  et~al\mbox{.}}{2015}]%
        {Bailey2015}
\bibfield{author}{\bibinfo{person}{Peter Bailey}, \bibinfo{person}{Alistair
  Moffat}, \bibinfo{person}{Falk Scholer}, {and} \bibinfo{person}{Paul
  Thomas}.} \bibinfo{year}{2015}\natexlab{}.
\newblock \showarticletitle{User Variability and IR System Evaluation}. In
  \bibinfo{booktitle}{\emph{Proceedings of the 38th International ACM SIGIR
  Conference on Research and Development in Information Retrieval}} (Santiago,
  Chile) \emph{(\bibinfo{series}{SIGIR '15})}. \bibinfo{publisher}{Association
  for Computing Machinery}, \bibinfo{address}{New York, NY, USA},
  \bibinfo{pages}{625–634}.
\newblock
\showISBNx{9781450336215}
\urldef\tempurl%
\url{https://doi.org/10.1145/2766462.2767728}
\showDOI{\tempurl}


\bibitem[\protect\citeauthoryear{Bogers, G{\"a}de, Koolen, Petras, and
  Skov}{Bogers et~al\mbox{.}}{2018}]%
        {bogers2019what}
\bibfield{author}{\bibinfo{person}{Toine Bogers}, \bibinfo{person}{Maria
  G{\"a}de}, \bibinfo{person}{Marijn Koolen}, \bibinfo{person}{Vivien Petras},
  {and} \bibinfo{person}{Mette Skov}.} \bibinfo{year}{2018}\natexlab{}.
\newblock \showarticletitle{``What was this Movie About this Chick?''}. In
  \bibinfo{booktitle}{\emph{Transforming Digital Worlds}},
  \bibfield{editor}{\bibinfo{person}{Gobinda Chowdhury}, \bibinfo{person}{Julie
  McLeod}, \bibinfo{person}{Val Gillet}, {and} \bibinfo{person}{Peter Willett}}
  (Eds.). \bibinfo{publisher}{Springer International Publishing},
  \bibinfo{address}{Cham}, \bibinfo{pages}{323--334}.
\newblock
\showISBNx{978-3-319-78105-1}


\bibitem[\protect\citeauthoryear{Broder}{Broder}{2002}]%
        {broder2002}
\bibfield{author}{\bibinfo{person}{Andrei Broder}.}
  \bibinfo{year}{2002}\natexlab{}.
\newblock \showarticletitle{A Taxonomy of Web Search}.
\newblock \bibinfo{journal}{\emph{SIGIR Forum}} \bibinfo{volume}{36},
  \bibinfo{number}{2} (\bibinfo{date}{Sept.} \bibinfo{year}{2002}),
  \bibinfo{pages}{3–10}.
\newblock
\showISSN{0163-5840}
\urldef\tempurl%
\url{https://doi.org/10.1145/792550.792552}
\showDOI{\tempurl}


\bibitem[\protect\citeauthoryear{El~Asri, Schulz, Sarma, Zumer, Harris, Fine,
  Mehrotra, and Suleman}{El~Asri et~al\mbox{.}}{2017}]%
        {el2017frames}
\bibfield{author}{\bibinfo{person}{Layla El~Asri}, \bibinfo{person}{Hannes
  Schulz}, \bibinfo{person}{Shikhar~Kr Sarma}, \bibinfo{person}{Jeremie Zumer},
  \bibinfo{person}{Justin Harris}, \bibinfo{person}{Emery Fine},
  \bibinfo{person}{Rahul Mehrotra}, {and} \bibinfo{person}{Kaheer Suleman}.}
  \bibinfo{year}{2017}\natexlab{}.
\newblock \showarticletitle{Frames: a corpus for adding memory to goal-oriented
  dialogue systems}. In \bibinfo{booktitle}{\emph{Proceedings of the 18th
  Annual SIGdial Meeting on Discourse and Dialogue}}.
  \bibinfo{pages}{207--219}.
\newblock


\bibitem[\protect\citeauthoryear{Furnas, Landauer, Gomez, and Dumais}{Furnas
  et~al\mbox{.}}{1987}]%
        {furnas1987vocabulary}
\bibfield{author}{\bibinfo{person}{George~W. Furnas},
  \bibinfo{person}{Thomas~K. Landauer}, \bibinfo{person}{Louis~M. Gomez}, {and}
  \bibinfo{person}{Susan~T. Dumais}.} \bibinfo{year}{1987}\natexlab{}.
\newblock \showarticletitle{The vocabulary problem in human-system
  communication}.
\newblock \bibinfo{journal}{\emph{Commun. ACM}} \bibinfo{volume}{30},
  \bibinfo{number}{11} (\bibinfo{year}{1987}), \bibinfo{pages}{964--971}.
\newblock


\bibitem[\protect\citeauthoryear{Guo, Xu, Cheng, and Li}{Guo
  et~al\mbox{.}}{2009}]%
        {jiafeng2009named}
\bibfield{author}{\bibinfo{person}{Jiafeng Guo}, \bibinfo{person}{Gu Xu},
  \bibinfo{person}{Xueqi Cheng}, {and} \bibinfo{person}{Hang Li}.}
  \bibinfo{year}{2009}\natexlab{}.
\newblock \showarticletitle{Named Entity Recognition in Query}. In
  \bibinfo{booktitle}{\emph{Proceedings of the 32nd International ACM SIGIR
  Conference on Research and Development in Information Retrieval}} (Boston,
  MA, USA) \emph{(\bibinfo{series}{SIGIR '09})}.
  \bibinfo{publisher}{Association for Computing Machinery},
  \bibinfo{address}{New York, NY, USA}, \bibinfo{pages}{267–274}.
\newblock
\showISBNx{9781605584836}
\urldef\tempurl%
\url{https://doi.org/10.1145/1571941.1571989}
\showDOI{\tempurl}


\bibitem[\protect\citeauthoryear{Guy}{Guy}{2016}]%
        {guy2016searching}
\bibfield{author}{\bibinfo{person}{Ido Guy}.} \bibinfo{year}{2016}\natexlab{}.
\newblock \showarticletitle{Searching by Talking: Analysis of Voice Queries on
  Mobile Web Search}. In \bibinfo{booktitle}{\emph{Proceedings of the 39th
  International ACM SIGIR Conference on Research and Development in Information
  Retrieval}} (Pisa, Italy) \emph{(\bibinfo{series}{SIGIR ’16})}.
  \bibinfo{publisher}{Association for Computing Machinery},
  \bibinfo{address}{New York, NY, USA}, \bibinfo{pages}{35–44}.
\newblock
\showISBNx{9781450340694}
\urldef\tempurl%
\url{https://doi.org/10.1145/2911451.2911525}
\showDOI{\tempurl}


\bibitem[\protect\citeauthoryear{Kammerer and Bohnacker}{Kammerer and
  Bohnacker}{2012}]%
        {kammerer2012children}
\bibfield{author}{\bibinfo{person}{Yvonne Kammerer} {and} \bibinfo{person}{Maja
  Bohnacker}.} \bibinfo{year}{2012}\natexlab{}.
\newblock \showarticletitle{Children's Web Search with Google: The
  Effectiveness of Natural Language Queries}. In
  \bibinfo{booktitle}{\emph{Proceedings of the 11th International Conference on
  Interaction Design and Children}} (Bremen, Germany)
  \emph{(\bibinfo{series}{IDC ’12})}. \bibinfo{publisher}{Association for
  Computing Machinery}, \bibinfo{address}{New York, NY, USA},
  \bibinfo{pages}{184–187}.
\newblock
\showISBNx{9781450310079}
\urldef\tempurl%
\url{https://doi.org/10.1145/2307096.2307121}
\showDOI{\tempurl}


\bibitem[\protect\citeauthoryear{Kato, Yamamoto, Ohshima, and Tanaka}{Kato
  et~al\mbox{.}}{2014}]%
        {kato2014cognitive}
\bibfield{author}{\bibinfo{person}{Makoto~P. Kato}, \bibinfo{person}{Takehiro
  Yamamoto}, \bibinfo{person}{Hiroaki Ohshima}, {and} \bibinfo{person}{Katsumi
  Tanaka}.} \bibinfo{year}{2014}\natexlab{}.
\newblock \showarticletitle{Cognitive Search Intents Hidden behind Queries: A
  User Study on Query Formulations}. In \bibinfo{booktitle}{\emph{Proceedings
  of the 23rd International Conference on World Wide Web}} (Seoul, Korea)
  \emph{(\bibinfo{series}{WWW ’14 Companion})}.
  \bibinfo{publisher}{Association for Computing Machinery},
  \bibinfo{address}{New York, NY, USA}, \bibinfo{pages}{313–314}.
\newblock
\showISBNx{9781450327459}
\urldef\tempurl%
\url{https://doi.org/10.1145/2567948.2577279}
\showDOI{\tempurl}


\bibitem[\protect\citeauthoryear{Koolen, Bogers, G{\"a}de, Hall, Hendrickx,
  Huurdeman, Kamps, Skov, Verberne, and Walsh}{Koolen et~al\mbox{.}}{2016}]%
        {koolen2016overview}
\bibfield{author}{\bibinfo{person}{Marijn Koolen}, \bibinfo{person}{Toine
  Bogers}, \bibinfo{person}{Maria G{\"a}de}, \bibinfo{person}{Mark Hall},
  \bibinfo{person}{Iris Hendrickx}, \bibinfo{person}{Hugo Huurdeman},
  \bibinfo{person}{Jaap Kamps}, \bibinfo{person}{Mette Skov},
  \bibinfo{person}{Suzan Verberne}, {and} \bibinfo{person}{David Walsh}.}
  \bibinfo{year}{2016}\natexlab{}.
\newblock \showarticletitle{Overview of the CLEF 2016 social book search lab}.
  In \bibinfo{booktitle}{\emph{International conference of the cross-language
  evaluation forum for European languages}}. Springer,
  \bibinfo{pages}{351--370}.
\newblock


\bibitem[\protect\citeauthoryear{Moraes, Yang, Zhang, and Murdock}{Moraes
  et~al\mbox{.}}{2020}]%
        {Moraes20}
\bibfield{author}{\bibinfo{person}{Felipe Moraes}, \bibinfo{person}{Jie Yang},
  \bibinfo{person}{Rongting Zhang}, {and} \bibinfo{person}{Vanessa Murdock}.}
  \bibinfo{year}{2020}\natexlab{}.
\newblock \showarticletitle{The Role of Attributes in Product Quality
  Comparisons}. In \bibinfo{booktitle}{\emph{Proceedings of the 2020 Conference
  on Human Information Interaction and Retrieval}} (Vancouver BC, Canada)
  \emph{(\bibinfo{series}{CHIIR '20})}. \bibinfo{publisher}{Association for
  Computing Machinery}, \bibinfo{address}{New York, NY, USA},
  \bibinfo{pages}{253–262}.
\newblock
\showISBNx{9781450368926}
\urldef\tempurl%
\url{https://doi.org/10.1145/3343413.3377956}
\showDOI{\tempurl}


\bibitem[\protect\citeauthoryear{of~Commerce}{of~Commerce}{2020}]%
        {uscommerce2020}
\bibfield{author}{\bibinfo{person}{U.S.~Department of Commerce}.}
  \bibinfo{year}{2020}\natexlab{}.
\newblock \bibinfo{booktitle}{\emph{{Quarterly Retail E-Commerce Sales}}}.
\newblock \bibinfo{type}{{T}echnical {R}eport}. \bibinfo{institution}{U.S.
  Department of Commerce}.
\newblock
\urldef\tempurl%
\url{https://www.census.gov/retail/mrts/www/data/pdf/ec_current.pdf}
\showURL{%
\tempurl}
\newblock
\shownote{Accessed on 22.10.2020.}


\bibitem[\protect\citeauthoryear{Papenmeier, Sliwa, Kern, Hienert, Aker, and
  Fuhr}{Papenmeier et~al\mbox{.}}{2020}]%
        {papenmeier2020modern}
\bibfield{author}{\bibinfo{person}{Andrea Papenmeier}, \bibinfo{person}{Alfred
  Sliwa}, \bibinfo{person}{Dagmar Kern}, \bibinfo{person}{Daniel Hienert},
  \bibinfo{person}{Ahmet Aker}, {and} \bibinfo{person}{Norbert Fuhr}.}
  \bibinfo{year}{2020}\natexlab{}.
\newblock \showarticletitle{'A Modern Up-To-Date Laptop'-Vagueness in Natural
  Language Queries for Product Search}. In
  \bibinfo{booktitle}{\emph{Proceedings of the 2020 ACM Designing Interactive
  Systems Conference}}. \bibinfo{pages}{2077--2089}.
\newblock


\bibitem[\protect\citeauthoryear{Penha, Balan, and Hauff}{Penha
  et~al\mbox{.}}{2019}]%
        {penha2019introducing}
\bibfield{author}{\bibinfo{person}{Gustavo Penha}, \bibinfo{person}{Alexandru
  Balan}, {and} \bibinfo{person}{Claudia Hauff}.}
  \bibinfo{year}{2019}\natexlab{}.
\newblock \showarticletitle{Introducing MANtIS: a novel multi-domain
  information seeking dialogues dataset}.
\newblock \bibinfo{journal}{\emph{arXiv preprint arXiv:1912.04639}}
  (\bibinfo{year}{2019}).
\newblock


\bibitem[\protect\citeauthoryear{Potthast, Hagen, V{\"o}lske, and
  Stein}{Potthast et~al\mbox{.}}{[n.d.]}]%
        {potthast2013exploratory}
\bibfield{author}{\bibinfo{person}{Martin Potthast}, \bibinfo{person}{Matthias
  Hagen}, \bibinfo{person}{Michael V{\"o}lske}, {and} \bibinfo{person}{Benno
  Stein}.} \bibinfo{year}{[n.d.]}\natexlab{}.
\newblock \showarticletitle{Exploratory Search Missions for TREC Topics.}
\newblock  (\bibinfo{year}{[n.\,d.]}).
\newblock


\bibitem[\protect\citeauthoryear{Qu, Yang, Croft, Trippas, Zhang, and Qiu}{Qu
  et~al\mbox{.}}{2018}]%
        {Qu2018}
\bibfield{author}{\bibinfo{person}{Chen Qu}, \bibinfo{person}{Liu Yang},
  \bibinfo{person}{W.~Bruce Croft}, \bibinfo{person}{Johanne~R. Trippas},
  \bibinfo{person}{Yongfeng Zhang}, {and} \bibinfo{person}{Minghui Qiu}.}
  \bibinfo{year}{2018}\natexlab{}.
\newblock \showarticletitle{Analyzing and Characterizing User Intent in
  Information-Seeking Conversations} \emph{(\bibinfo{series}{SIGIR '18})}.
  \bibinfo{publisher}{Association for Computing Machinery},
  \bibinfo{address}{New York, NY, USA}, \bibinfo{pages}{989–992}.
\newblock
\showISBNx{9781450356572}
\urldef\tempurl%
\url{https://doi.org/10.1145/3209978.3210124}
\showDOI{\tempurl}


\bibitem[\protect\citeauthoryear{Radlinski, Balog, Byrne, and
  Krishnamoorthi}{Radlinski et~al\mbox{.}}{2019}]%
        {Radlinski2019}
\bibfield{author}{\bibinfo{person}{Filip Radlinski}, \bibinfo{person}{Krisztian
  Balog}, \bibinfo{person}{Bill Byrne}, {and} \bibinfo{person}{Karthik
  Krishnamoorthi}.} \bibinfo{year}{2019}\natexlab{}.
\newblock \showarticletitle{Coached Conversational Preference Elicitation: A
  Case Study in Understanding Movie Preferences}. In
  \bibinfo{booktitle}{\emph{Proceedings of the Annual SIGdial Meeting on
  Discourse and Dialogue}}.
\newblock


\bibitem[\protect\citeauthoryear{Sadikov, Madhavan, Wang, and Halevy}{Sadikov
  et~al\mbox{.}}{2010}]%
        {sadikov2010clustering}
\bibfield{author}{\bibinfo{person}{Eldar Sadikov}, \bibinfo{person}{Jayant
  Madhavan}, \bibinfo{person}{Lu Wang}, {and} \bibinfo{person}{Alon Halevy}.}
  \bibinfo{year}{2010}\natexlab{}.
\newblock \showarticletitle{Clustering Query Refinements by User Intent}. In
  \bibinfo{booktitle}{\emph{Proceedings of the 19th International Conference on
  World Wide Web}} (Raleigh, North Carolina, USA) \emph{(\bibinfo{series}{WWW
  '10})}. \bibinfo{publisher}{Association for Computing Machinery},
  \bibinfo{address}{New York, NY, USA}, \bibinfo{pages}{841–850}.
\newblock
\showISBNx{9781605587998}
\urldef\tempurl%
\url{https://doi.org/10.1145/1772690.1772776}
\showDOI{\tempurl}


\bibitem[\protect\citeauthoryear{Shen, Yan, Yan, Ji, Liu, and Chen}{Shen
  et~al\mbox{.}}{2011}]%
        {shen2011sparse}
\bibfield{author}{\bibinfo{person}{Yelong Shen}, \bibinfo{person}{Jun Yan},
  \bibinfo{person}{Shuicheng Yan}, \bibinfo{person}{Lei Ji},
  \bibinfo{person}{Ning Liu}, {and} \bibinfo{person}{Zheng Chen}.}
  \bibinfo{year}{2011}\natexlab{}.
\newblock \showarticletitle{Sparse Hidden-Dynamics Conditional Random Fields
  for User Intent Understanding}. In \bibinfo{booktitle}{\emph{Proceedings of
  the 20th International Conference on World Wide Web}} (Hyderabad, India)
  \emph{(\bibinfo{series}{WWW '11})}. \bibinfo{publisher}{Association for
  Computing Machinery}, \bibinfo{address}{New York, NY, USA},
  \bibinfo{pages}{7–16}.
\newblock
\showISBNx{9781450306324}
\urldef\tempurl%
\url{https://doi.org/10.1145/1963405.1963411}
\showDOI{\tempurl}


\bibitem[\protect\citeauthoryear{Skinner and Kallumadi}{Skinner and
  Kallumadi}{2019}]%
        {skinner2019commerce}
\bibfield{author}{\bibinfo{person}{Michael Skinner} {and}
  \bibinfo{person}{Surya Kallumadi}.} \bibinfo{year}{2019}\natexlab{}.
\newblock \showarticletitle{E-commerce Query Classification Using Product
  Taxonomy Mapping: A Transfer Learning Approach.}. In
  \bibinfo{booktitle}{\emph{eCOM@ SIGIR}}.
\newblock


\bibitem[\protect\citeauthoryear{Sondhi, Sharma, Kolari, and Zhai}{Sondhi
  et~al\mbox{.}}{2018}]%
        {Sondhi2018}
\bibfield{author}{\bibinfo{person}{Parikshit Sondhi}, \bibinfo{person}{Mohit
  Sharma}, \bibinfo{person}{Pranam Kolari}, {and} \bibinfo{person}{ChengXiang
  Zhai}.} \bibinfo{year}{2018}\natexlab{}.
\newblock \showarticletitle{A Taxonomy of Queries for E-Commerce Search}. In
  \bibinfo{booktitle}{\emph{The 41st International ACM SIGIR Conference on
  Research \& Development in Information Retrieval}} (Ann Arbor, MI, USA)
  \emph{(\bibinfo{series}{SIGIR '18})}. \bibinfo{publisher}{Association for
  Computing Machinery}, \bibinfo{address}{New York, NY, USA},
  \bibinfo{pages}{1245–1248}.
\newblock
\showISBNx{9781450356572}
\urldef\tempurl%
\url{https://doi.org/10.1145/3209978.3210152}
\showDOI{\tempurl}


\bibitem[\protect\citeauthoryear{Sorokina and Cantu-Paz}{Sorokina and
  Cantu-Paz}{2016}]%
        {sorokina2016amazon}
\bibfield{author}{\bibinfo{person}{Daria Sorokina} {and} \bibinfo{person}{Erick
  Cantu-Paz}.} \bibinfo{year}{2016}\natexlab{}.
\newblock \showarticletitle{Amazon search: The joy of ranking products}. In
  \bibinfo{booktitle}{\emph{Proceedings of the 39th International ACM SIGIR
  conference on Research and Development in Information Retrieval}}.
  \bibinfo{pages}{459--460}.
\newblock


\bibitem[\protect\citeauthoryear{Su, He, Liu, Zhang, and Ma}{Su
  et~al\mbox{.}}{2018}]%
        {su2018}
\bibfield{author}{\bibinfo{person}{Ning Su}, \bibinfo{person}{Jiyin He},
  \bibinfo{person}{Yiqun Liu}, \bibinfo{person}{Min Zhang}, {and}
  \bibinfo{person}{Shaoping Ma}.} \bibinfo{year}{2018}\natexlab{}.
\newblock \showarticletitle{User Intent, Behaviour, and Perceived Satisfaction
  in Product Search}. In \bibinfo{booktitle}{\emph{Proceedings of the Eleventh
  ACM International Conference on Web Search and Data Mining}} (Marina Del Rey,
  CA, USA) \emph{(\bibinfo{series}{WSDM '18})}. \bibinfo{publisher}{Association
  for Computing Machinery}, \bibinfo{address}{New York, NY, USA},
  \bibinfo{pages}{547–555}.
\newblock
\showISBNx{9781450355810}
\urldef\tempurl%
\url{https://doi.org/10.1145/3159652.3159714}
\showDOI{\tempurl}


\bibitem[\protect\citeauthoryear{Thomas, McDuff, Czerwinski, and
  Craswell}{Thomas et~al\mbox{.}}{2017}]%
        {thomas2017misc}
\bibfield{author}{\bibinfo{person}{Paul Thomas}, \bibinfo{person}{Daniel
  McDuff}, \bibinfo{person}{Mary Czerwinski}, {and} \bibinfo{person}{Nick
  Craswell}.} \bibinfo{year}{2017}\natexlab{}.
\newblock \showarticletitle{MISC: A data set of information-seeking
  conversations}. In \bibinfo{booktitle}{\emph{SIGIR 1st International Workshop
  on Conversational Approaches to Information Retrieval (CAIR’17)}},
  Vol.~\bibinfo{volume}{5}.
\newblock


\bibitem[\protect\citeauthoryear{Trippas, Spina, Cavedon, and
  Sanderson}{Trippas et~al\mbox{.}}{2017}]%
        {trippas2017how}
\bibfield{author}{\bibinfo{person}{Johanne~R. Trippas},
  \bibinfo{person}{Damiano Spina}, \bibinfo{person}{Lawrence Cavedon}, {and}
  \bibinfo{person}{Mark Sanderson}.} \bibinfo{year}{2017}\natexlab{}.
\newblock \showarticletitle{How Do People Interact in Conversational
  Speech-Only Search Tasks: A Preliminary Analysis}. In
  \bibinfo{booktitle}{\emph{Proceedings of the 2017 Conference on Conference
  Human Information Interaction and Retrieval}} (Oslo, Norway)
  \emph{(\bibinfo{series}{CHIIR '17})}. \bibinfo{publisher}{ACM},
  \bibinfo{address}{New York, NY, USA}, \bibinfo{pages}{325--328}.
\newblock
\urldef\tempurl%
\url{https://doi.org/10.1145/3020165.3022144}
\showDOI{\tempurl}


\bibitem[\protect\citeauthoryear{Trotman, Degenhardt, and Kallumadi}{Trotman
  et~al\mbox{.}}{2017}]%
        {Trotman2017}
\bibfield{author}{\bibinfo{person}{Andrew Trotman}, \bibinfo{person}{Jon
  Degenhardt}, {and} \bibinfo{person}{Surya Kallumadi}.}
  \bibinfo{year}{2017}\natexlab{}.
\newblock \showarticletitle{The Architecture of eBay Search}. In
  \bibinfo{booktitle}{\emph{Proceedings of the {SIGIR} 2017 Workshop On
  eCommerce co-located with the 40th International {ACM} {SIGIR} Conference on
  Research and Development in Information Retrieval, eCOM@SIGIR 2017, Tokyo,
  Japan, August 11, 2017}} \emph{(\bibinfo{series}{{CEUR} Workshop
  Proceedings}, Vol.~\bibinfo{volume}{2311})},
  \bibfield{editor}{\bibinfo{person}{Jon Degenhardt}, \bibinfo{person}{Surya
  Kallumadi}, \bibinfo{person}{Maarten de~Rijke}, \bibinfo{person}{Luo Si},
  \bibinfo{person}{Andrew Trotman}, {and} \bibinfo{person}{Yinghui Xu}} (Eds.).
  \bibinfo{publisher}{CEUR-WS.org}.
\newblock
\urldef\tempurl%
\url{http://ceur-ws.org/Vol-2311/paper\_14.pdf}
\showURL{%
\tempurl}


\bibitem[\protect\citeauthoryear{Tsagkias, King, Kallumadi, Murdock, and
  de~Rijke}{Tsagkias et~al\mbox{.}}{2020}]%
        {tsagkias2020challenges}
\bibfield{author}{\bibinfo{person}{Manos Tsagkias},
  \bibinfo{person}{Tracy~Holloway King}, \bibinfo{person}{Surya Kallumadi},
  \bibinfo{person}{Vanessa Murdock}, {and} \bibinfo{person}{Maarten de Rijke}.}
  \bibinfo{year}{2020}\natexlab{}.
\newblock \showarticletitle{Challenges and Research Opportunities in eCommerce
  Search and Recommendations}. In \bibinfo{booktitle}{\emph{SIGIR Forum}},
  Vol.~\bibinfo{volume}{54}.
\newblock


\bibitem[\protect\citeauthoryear{Voorhees and Harman}{Voorhees and
  Harman}{2005}]%
        {vorhees2005}
\bibfield{author}{\bibinfo{person}{Ellen~M. Voorhees} {and}
  \bibinfo{person}{Donna~K. Harman}.} \bibinfo{year}{2005}\natexlab{}.
\newblock \bibinfo{booktitle}{\emph{TREC: Experiment and Evaluation in
  Information Retrieval (Digital Libraries and Electronic Publishing)}}.
\newblock \bibinfo{publisher}{The MIT Press}.
\newblock
\showISBNx{0262220733}


\bibitem[\protect\citeauthoryear{Wang and Zhao}{Wang and Zhao}{2019}]%
        {wang2019multi}
\bibfield{author}{\bibinfo{person}{Chao Wang} {and} \bibinfo{person}{Rongkai
  Zhao}.} \bibinfo{year}{2019}\natexlab{}.
\newblock \showarticletitle{Multi-Candidate Ranking Algorithm Based Spell
  Correction.}. In \bibinfo{booktitle}{\emph{Proceedings of the {SIGIR} 2019
  Workshop On eCommerce co-located with the 42st International ACM SIGIR
  Conference on Research and Development in Information Retrieval (SIGIR 2019)
  Paris, France, July 25, 2019.}} \emph{(\bibinfo{series}{{CEUR} Workshop
  Proceedings})}, \bibfield{editor}{\bibinfo{person}{Jon Degenhardt},
  \bibinfo{person}{Surya Kallumadi~Utkarsh Porwal}, {and}
  \bibinfo{person}{Andrew Trotman}} (Eds.).
\newblock


\bibitem[\protect\citeauthoryear{Wang, Huang, Zhao, Zhang, Zhao, and Lee}{Wang
  et~al\mbox{.}}{2018}]%
        {wang2018billion}
\bibfield{author}{\bibinfo{person}{Jizhe Wang}, \bibinfo{person}{Pipei Huang},
  \bibinfo{person}{Huan Zhao}, \bibinfo{person}{Zhibo Zhang},
  \bibinfo{person}{Binqiang Zhao}, {and} \bibinfo{person}{Dik~Lun Lee}.}
  \bibinfo{year}{2018}\natexlab{}.
\newblock \showarticletitle{Billion-scale commodity embedding for e-commerce
  recommendation in alibaba}. In \bibinfo{booktitle}{\emph{Proceedings of the
  24th ACM SIGKDD International Conference on Knowledge Discovery \& Data
  Mining}}. \bibinfo{pages}{839--848}.
\newblock


\bibitem[\protect\citeauthoryear{Yan, Duan, Chen, Zhou, Zhou, and Li}{Yan
  et~al\mbox{.}}{2017}]%
        {yan2017building}
\bibfield{author}{\bibinfo{person}{Zhao Yan}, \bibinfo{person}{Nan Duan},
  \bibinfo{person}{Peng Chen}, \bibinfo{person}{Ming Zhou},
  \bibinfo{person}{Jianshe Zhou}, {and} \bibinfo{person}{Zhoujun Li}.}
  \bibinfo{year}{2017}\natexlab{}.
\newblock \showarticletitle{Building Task-Oriented Dialogue Systems for Online
  Shopping}. In \bibinfo{booktitle}{\emph{Proceedings of the Thirty-First AAAI
  Conference on Artificial Intelligence}} (San Francisco, California, USA)
  \emph{(\bibinfo{series}{AAAI'17})}. \bibinfo{publisher}{AAAI Press},
  \bibinfo{pages}{4618–4625}.
\newblock


\bibitem[\protect\citeauthoryear{Yu and Litchfield}{Yu and Litchfield}{2020}]%
        {yu2020query}
\bibfield{author}{\bibinfo{person}{Hang Yu} {and} \bibinfo{person}{Lester
  Litchfield}.} \bibinfo{year}{2020}\natexlab{}.
\newblock \showarticletitle{Query Classification with Multi-objective Backoff
  Optimization}. In \bibinfo{booktitle}{\emph{Proceedings of the 43rd
  International ACM SIGIR Conference on Research and Development in Information
  Retrieval}}. \bibinfo{pages}{1925--1928}.
\newblock


\end{thebibliography}

\end{document}